\documentclass[11pt]{article}
\usepackage{vietnam,epsfig}

\def\be{\begin{equation}}
\def\ee{\end{equation}}
\def\bea{\begin{eqnarray}}
\def\eea{\end{eqnarray}}

\begin{document}
\vspace*{4cm}
\title{Experimental tests of QCD at hadronic colliders}

\author{ C. Royon }

\address{DAPNIA-SPP, CE Saclay, F 91 191 Gif-sur-Yvette cedex, France}

\maketitle\abstracts{
We present a general overview of recent QCD results at hadronic colliders.}

\section{Proton structure function measurements}

\subsection{Structure function measurements at HERA}
The H1 and ZEUS experiments at HERA, at DESY, Hamburg, Germany allow to probe 
directly the quark and gluon contents of the proton by measuring $F_2$, the
proton structure function in bins of $x$, the proton momentum fraction
carried by the struck quark and $Q^2$, the squared energy transferred at the lepton
vertex. The advantage of an $ep$ collider is that different methods are
available to measure $x$ and $Q^2$ using either the lepton or the jets in the
final state or both. The kinematical plane reached by the HERA collaborations
is given in Fig. 1. We notice that the proton structure can be probed over a 
large kinematical plane ($10^{-6} < x$, $0.1 < Q^2 < 3. 10^4$), and has a common
domain with the previous fixed target experiments (SLAC, BCDMS, NMC...), which
allows a direct comparisons of the measurements.

The proton structure function $F_2$ measurement \cite{f2} is given in Fig. 2.
We notice the high precision of the measurement performed at HERA, in good 
agreement with the fixed target data. The lowest HERA $Q^2$ data benefit
from a good acceptance of the calorimeters at low $Q^2$ (SPACAL for H1, and 
Beam Pipe Calorimeter for ZEUS) and from special runs with shifted positions
of the interaction vertex which allow to measure lower angles for the scattered
electrons, and thus lower $Q^2$ events.

At moderate $Q^2$ and high $x$
($0.015 <x<0.65$, $1 < Q^2 <100$ GeV$^2$), the NuTeV collaboration 
\cite{nutev} has recently released
new measurements of the proton structure functions $F_2$ and $xF_3$ with high 
precision, in agreement with QCD expectations.

The high luminosity obtained at HERA allows to measure the neutral and charged
cross section with higher precision \cite{chargedcurrent}. A good
agreement is found between the standard model expectations and the measurements.

\subsection{DGLAP fits and extraction of the parton densities in the proton}
The $F_2$ data
allow to test precisely the Dokshitzer Gribov Lipatov Altarelli Parisi (DGLAP)
\cite{dglap} 
evolution equations and to obtain the quark and gluon contents in the proton at
NLO. In Fig. 2, we also see the good agreement between the measurement and 
the H1 and ZEUS DGLAP NLO fits. The parton distributions - quark and gluon
densities from the H1, ZEUS and CTEQ\cite{cteq} fits
\footnote{Results from MRS \cite{cteq} fits are similar, except for the gluon
at high $x$ which shows much difference}- are given in Fig. 3. The H1 and ZEUS 
collaborations use the data from their own experiments, together with
the data from the BCDMS and NMC experiments to constrain high-$x$ distributions
and valence quarks.
We notice the
high increase of the gluon density towards low values of $x$, which was first
discovered at HERA ten years ago. We also notice the good agreement between the different
fits. It would also be interesting to perform more global fits involving jet
cross section measurements, as the ZEUS collaboration started to do
\cite{mandy}.
\begin{figure}
\begin{center}
\epsfig{file=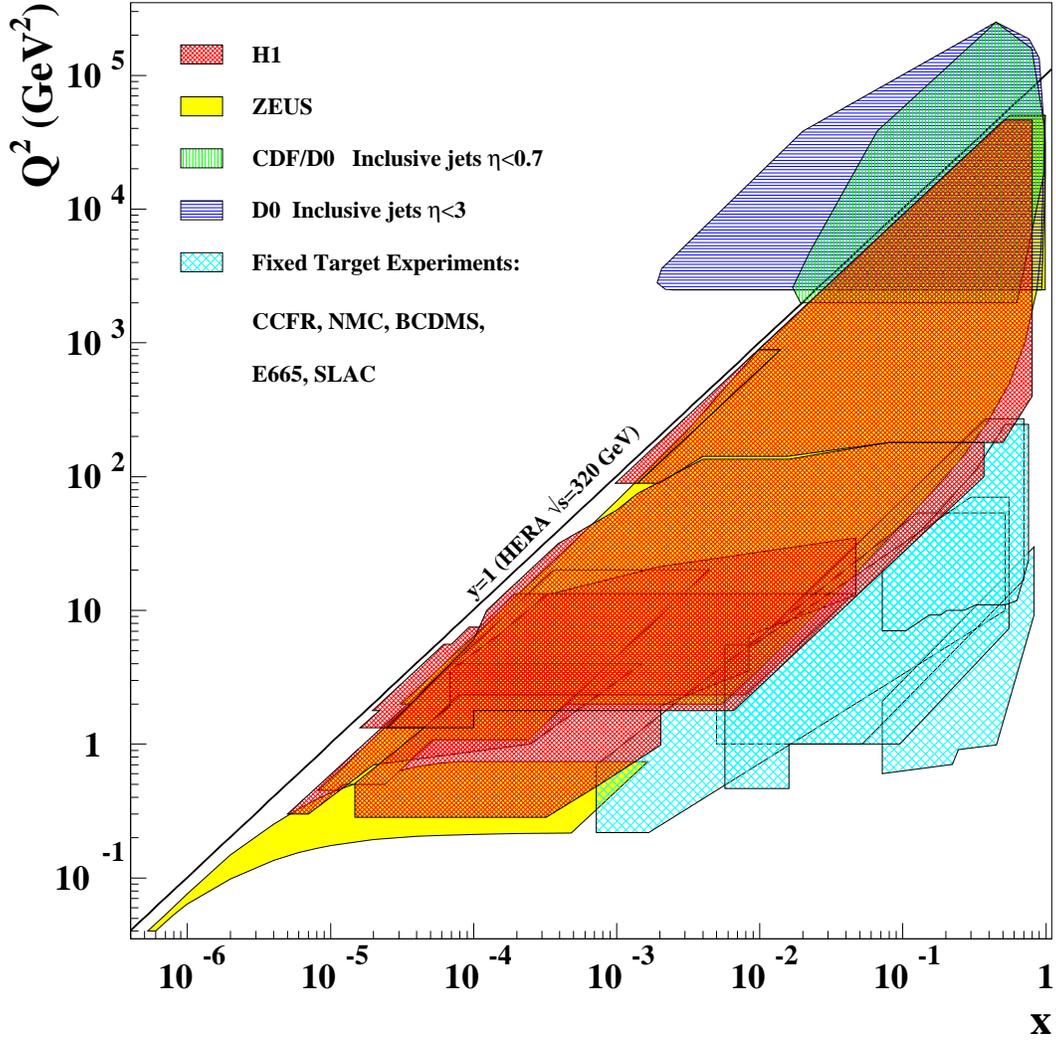,width=14cm}
\caption{HERA and Tevatron kinematical domains}
\end{center}
\label{kin}
\end{figure} 


\begin{figure}
\begin{center}
\vspace{20cm}
\epsfig{file=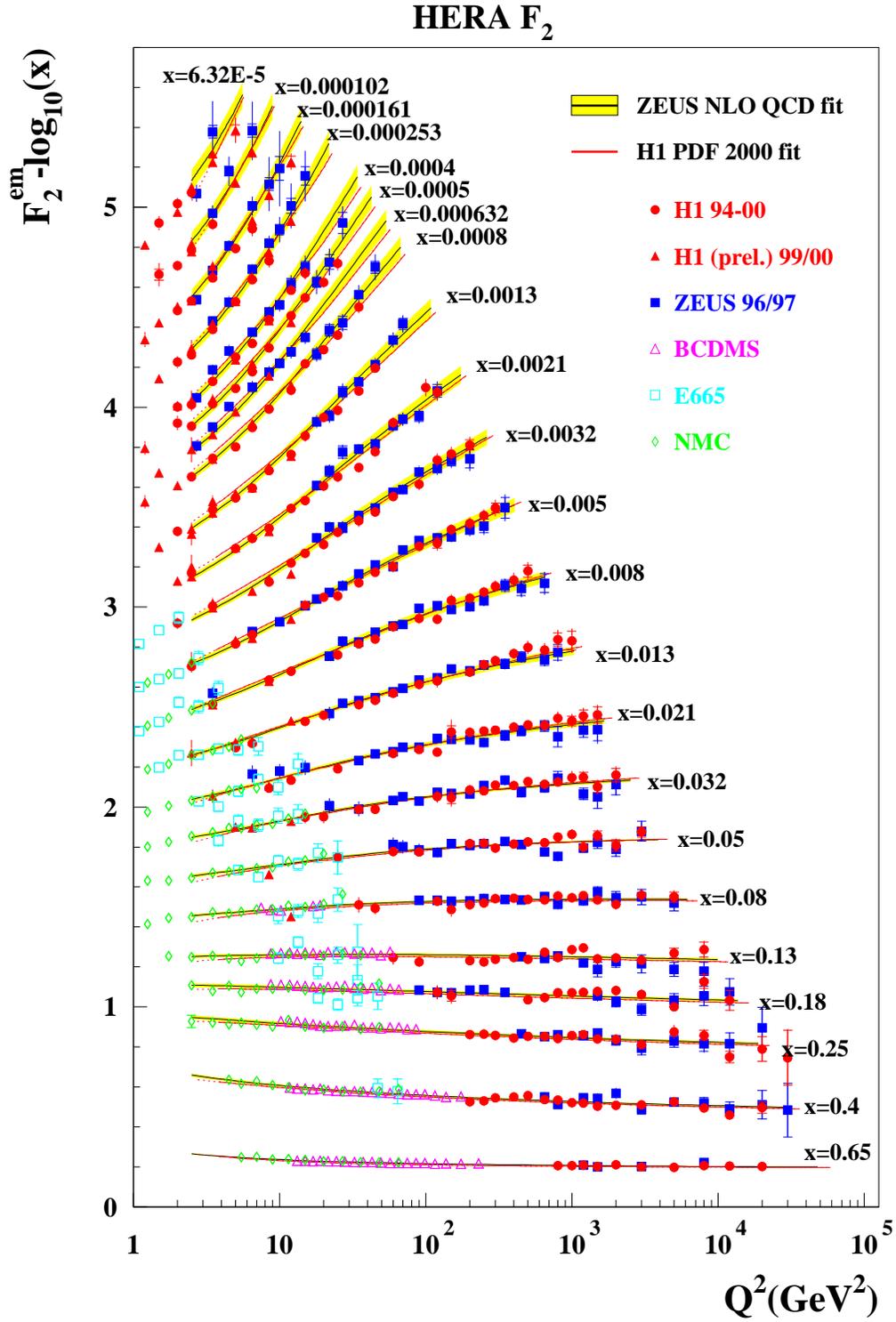,width=14cm, angle=180}
\caption{Proton structure function $F_2$ from the H1 and ZEUS experiments}
\end{center}
\label{f2}
\end{figure}

The gluon distribution at high $x$ is still badly known. The uncertainty
of $xG$ at $Q^2 = 100$ GeV$^2$, $x \sim$ 0.5, is of the order of 50\%. We will
see in the following that the complementarity between the HERA and Tevatron
measurements will allow to bring some knowledge on the high $x$ gluon density.

\begin{figure}[ht]
\begin{center}
\epsfig{file=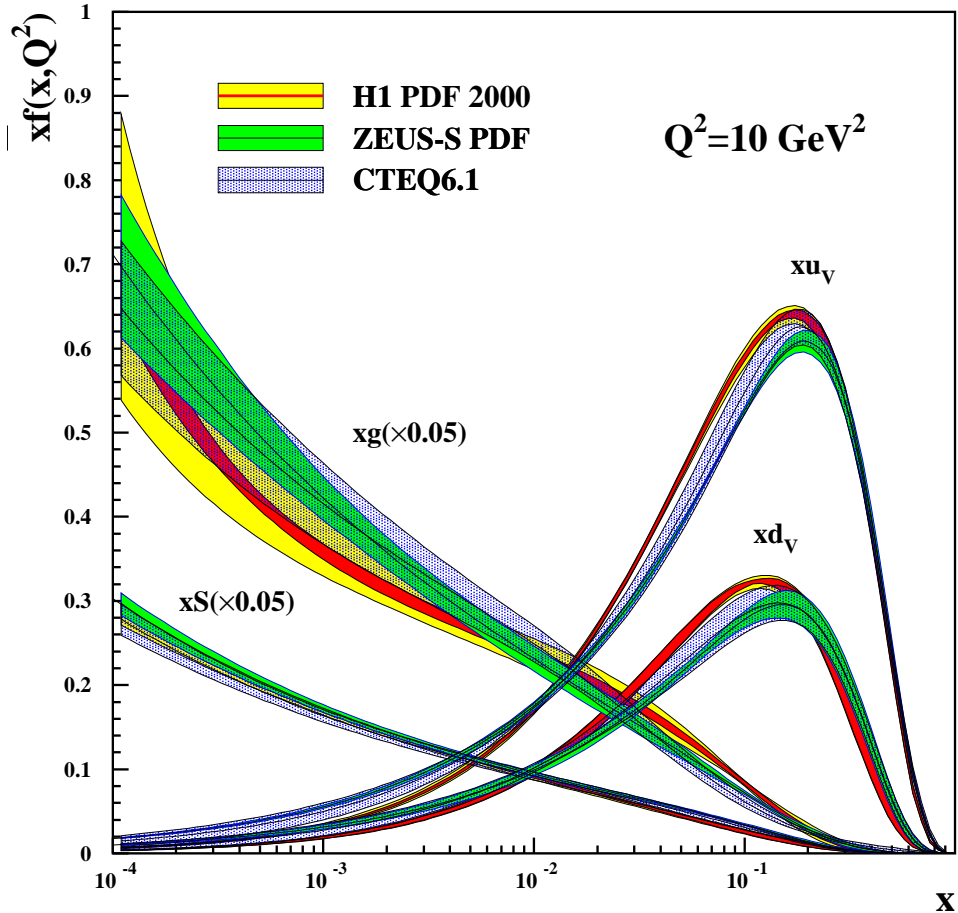,width=14cm}
\caption{Quark and gluon densities obtained from a NLO DGLAP fit to the proton 
structure function $F_2$}
\end{center}
\label{gluon}
\end{figure}

\subsection{$\alpha_S$ measurement}
The value of $\alpha_S (M_Z)$ comes directly as an output of the QCD DGLAP fits
of the proton structure functions. Other measurements performed at hadronic
colliders are made using inclusive jet cross section in $pp$ colliders, 
jet shapes, subjet multiplicities, inclusive jet cross section measurements
in $ep$ colliders. The different measurements, together with the references,
are given in Fig. 4. We notice that the experimental uncertainties on
$\alpha_S$ measurements at hadronic colliders approaches the uncertainties of
the LEP measurements (see the world average in Fig. 4), but the theoretical
uncertainties are much larger.

\begin{figure}[ht]
\begin{center}
\epsfig{file=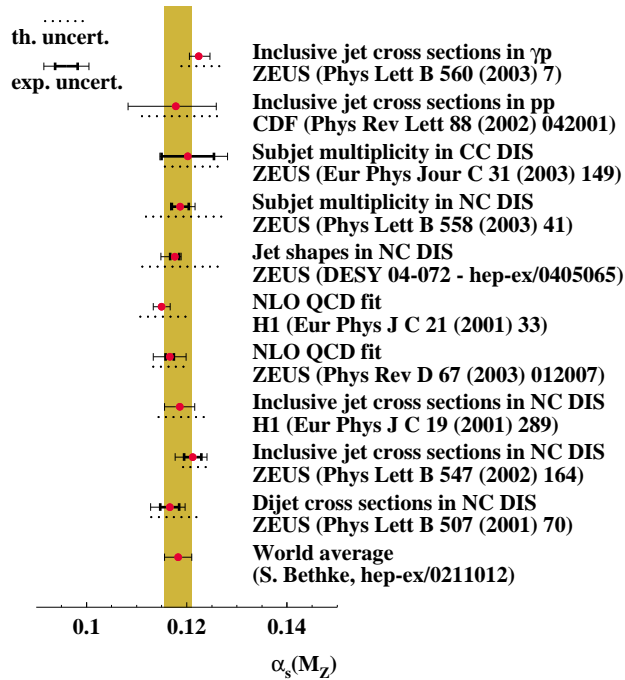,width=10cm}
\caption{$\alpha_S$ measurements performed at hadron colliders compared
to the world average.}
\end{center}
\label{alphaS}
\end{figure}

\section{Jet cross section measurements at the Tevatron}

As we mentioned earlier (see Fig. 1), the kinematical plane reached at the
Tevatron is complementary from the reach at HERA. We noticed that the high-$x$ gluon
density was purely constrained using HERA data, and the acceptance of the D\O\
and CDF experiments at high-$x$ will allow to obtain a better constraint of the
parton distributions in this kinematical domain. The Tevatron measurements are
specially important for searches at the LHC (multijet environment specially
for searches for $R$-parity violated SUSY), and the constraint of the 
high-$x$ gluon density (search for higher dimensions).

The inclusive jet cross section from the D\O\ collaboration
as a function of the transverse momentum
is shown in Fig. 5 (left) in different bins of rapidity
\cite{d0cdfjets}. This measurement is 
motivated by the run I excess observed by the CDF and D\O\ collaborations
at high jet $p_T$ which lead to a modification of the high-$x$ gluon 
density. The uncertainties, dominated
by the systematic errors on jet energy scale are still large especially at high
rapidity, and will be notably improved in the near future. The CDF collaboration
has similar results. Due to the size of the uncertainties, it is too early to
constrain the parton distributions - and especially, the gluon density at
high $x$ - using these data sets. The D\O\ collaboration has also performed
a measurement of the dijet cross section as a function of their mass.

The D\O\ collaboration has also measured the distribution in
difference in azimuthal angle $\Delta \Phi$  between
jets in multi-jet events \cite{deltaphi}
(see Fig. 5, right). The advantage of that measurement is that
it is sensitive to multijet events (3, 4, 5 jets...) without measuring 
the jet $p_T$ and is thus less sensitive to jet energy scale uncertainties.
Fig. 7 shows the good agreement between NLO QCD and the measurement except
at high values of $\Delta \Phi$ where the measurement is sensitive to soft
radiatio effects. At low $\Delta \Phi$, this measurement is also sensitive to
higher order calculation. The measurement also allows to tune the generators
like PYTHIA \cite{pythia}  because of its sensitivity on the amount of initial radiation
\cite{deltaphi} which is important for the LHC. The CDF collaboration
has also measured the underlying event properties by analyzing the charged
particle energy and multiplicity emitted in the region in azimuthal angle
outside the jets region in clean dijet events (see Ref. \cite{d0cdfjets}) which
is found to be in good agreement with PYTHIA expectations.
 
\begin{figure}
\begin{center}
\begin{tabular}{cc}
\hspace{-2cm}
\epsfig{figure=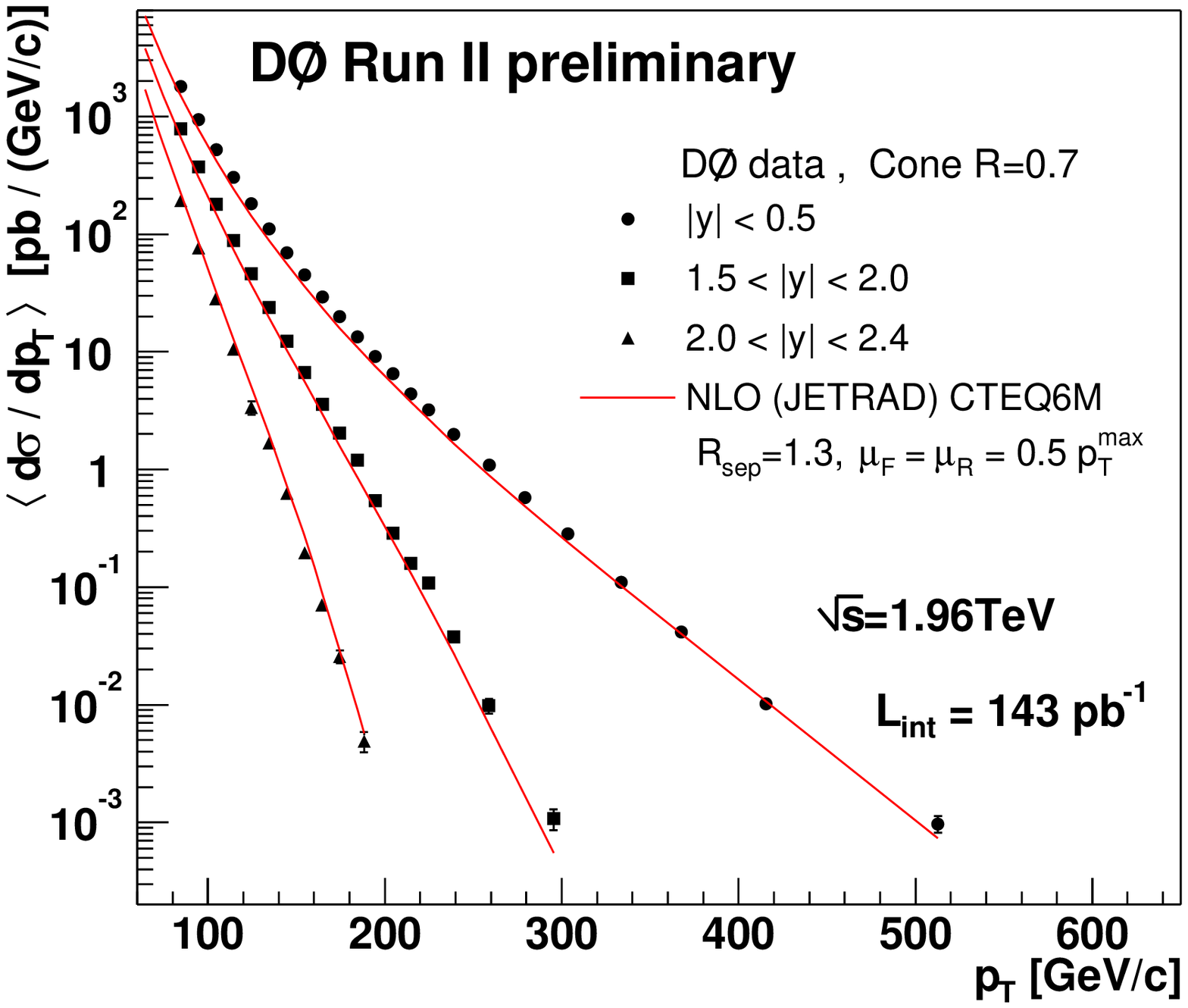,height=3.5in} &
\hspace{-0.8cm}
\epsfig{figure=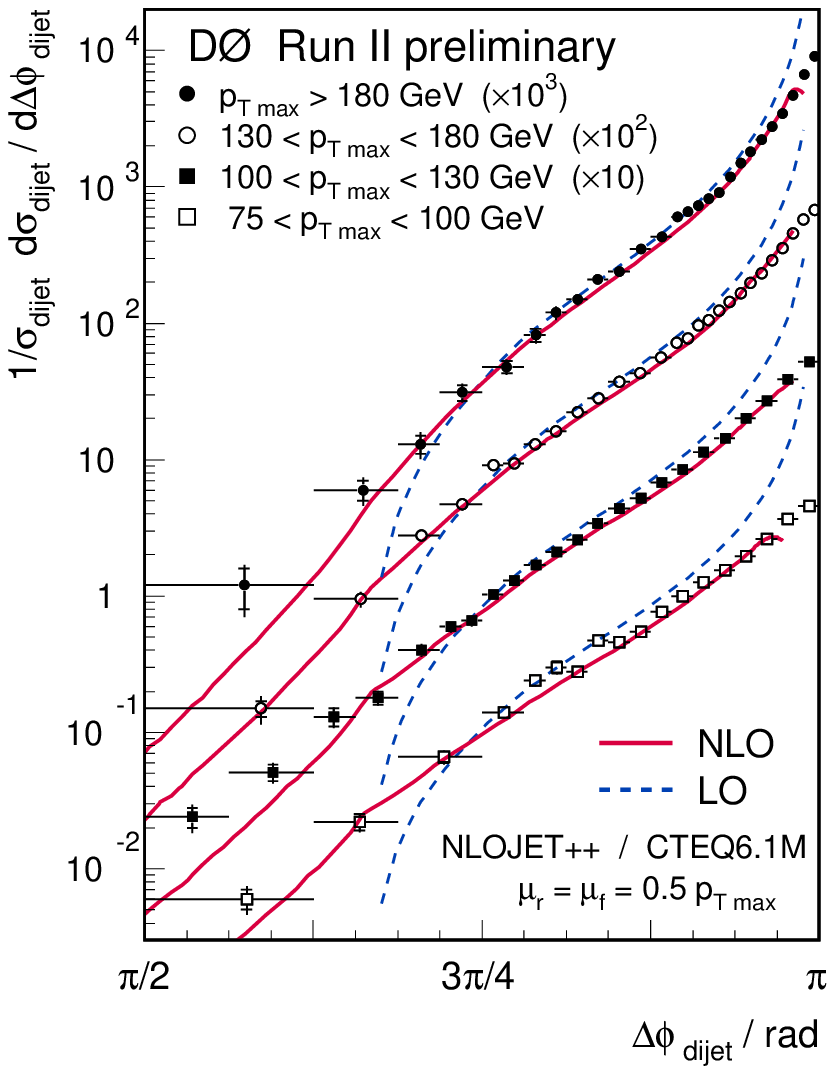,height=3.51in} \\
\end{tabular}
\caption{Inclusive $p_T$ jet cross section measurement and 
difference in azimuthal angles between jets at the Tevatron
(D\O\ experiment).}
\end{center}
\end{figure}

\section{BFKL resummation effects and saturation}
QCD evolution equations based on a resummation of $x$ terms have been proposed
some time ago by Baltiskii, Fadin, Kuraev and Lipatov first at LO and then at
NLO \cite{bfkl}. There is so far no evidence of the need of the BFKL resummation
at low $x$ in inclusive quantities such as the proton structure function $F_2$
even if some studies show the compatibility of the low x $F_2$ data and the 
BFKL resummation approach \cite{dipoles} and the BFKL effects can explain
the rise of the gluon towards low $x$. As a consequence,
BFKL resummation effects have been looked at in less inclusive quantities such
as the production of jets in the forward region at HERA or Mueller-Navelet jets
at the Tevatron. The idea is quite simple: one looks for jets with similar $k_T$
separated with large intervals in rapidity to suppress DGLAP evolution and
favour BFKL effects at the Tevatron or the LHC. In the same sense, at HERA,
it is possible to look for jets in the forward direction, far away from the
scattered electron, and with $p_T \sim Q^2$ of the virtual photon.

The H1 collaboration has measured the differential forward jet cross section
$d \sigma / d x d Q^2 d p_T^2$ in different bins of jet $p_T$, $x$, and
$Q^2$ (see Ref. \cite{forwardjets} and Fig. 6). The measurement is compared with NLO
DGLAP expectation and a clear excess is observed in the low $x$, low jet $p_T$
region. However, it is not obvious that this is a pure BFKL resummation effect
since the hadronisation corrections for low $p_T$ jets are high, and results
could be also interpreted as being due to the resolved component of the 
photon itself.

The forward jets at HERA and the Mueller-Navelet jets at the Tevatron, and 
at the LHC, can also be a nice measurement to look for saturation effects
\cite{cyrille}. An attempt to look for saturation at HERA using the forward jet
measurement was performed recently (see Ref. \cite{cyrille}) and leads to two
possible solutions with weak or large saturation effects. Mueller-Navelet
jet measurements at the LHC will allow to distinguish betweeen these two
solutions.  

\section{Conclusion}
In this short report, we presented many recent results concerning QCD at
hedronic colliders. The precision of the present inclusive cross section
measurements allow to obtain the parton densities in the proton with high
accuracy over a wide kinematical range. A measurement of the longitudinal
structure function at HERA by reducing the beam energies towards the end of
the HERA program would be of great importance to complete the determination of
the parton distributions and to test further the DGLAP evolution equations.
The Tevatron jet data allow to constrain further the gluon distribution
especially at high $x$ and some more measurements with higher precision are
expected in the near future. The $\alpha_S$ measurements at hadronic colliders
reach the experimental precision obtained at LEP, but the theoretical
uncertainties are still large. Searches for BFKL resummation and 
saturation effects have also been made at HERA especially in the forward jet
cross section measurements, and more measurements, especially at the  LHC
will be of great importance. Many other topics in QCD such as event shapes,
diffraction... have not been described in this talk because of the lack of time.

\begin{table}
\begin{center}
\epsfig{figure=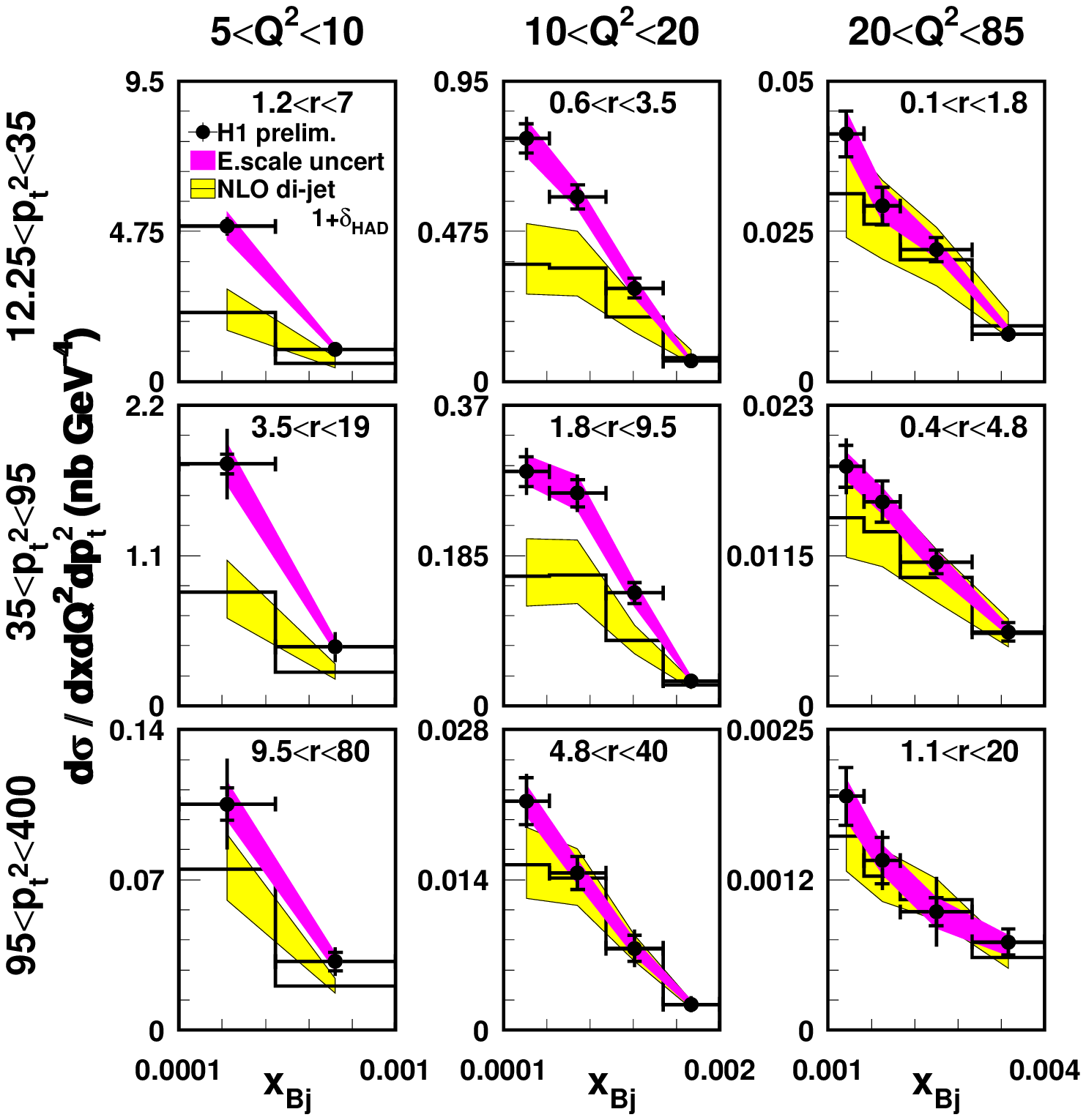,height=5in}
\caption{Saturation}
\end{center}
\end{table}

\end{document}